\documentclass[12pt]{book}
\usepackage{plenum}
\usepackage{epsfig}
\begin{document}

\setlength{\unitlength}{1mm}

\textheight 225mm
\def\sepand{\rule{14cm}{0pt}\and}

\flushbottom




\chapter{POTENTIAL TOPOGRAPHY AND MASS GENERATION}

\begin{center}

BCUNY-HEP-97-1
                         
\end{center}

\author{Maxime Kudinov\refnote{1,2}, Enrique Moreno\refnote{1,3}
and Peter Orland\refnote{1,4}}

\affiliation{\affnote{1}Baruch College, The City
University of New York\\ New York, NY 10010, USA\\
\affnote{2}Queens College, The City University
of New York\\ Flushing, NY 11367, USA\\
\affnote{3}The City College of the City
University of New York\\ New York, NY 10031, USA\\
\affnote{4}The Graduate School and University
Center\\ The City University of New York, New York, NY 10036, USA}

\vspace{20pt}




{\it Invited 
talk given by P. Orland at the NATO Advanced Workshop on Theoretical
Physics, ``New Developments in Quantum Field 
Theory," Zakopane, Poland, June 14-19, 1997, proceedings to be published
by Plenum Press.}

\section{INTRODUCTION}

\indent We describe an approach to understanding exponential decay
of correlation functions in asymptotically free theories. This
approach is systematic; it does not start from any 
conjectured mechanism or picture. We begin by studying\begin{itemize}
\item the 
metric on the space of configurations and
\item the behavior of the potential-energy function on this space.
\end{itemize}

We begin by describing how these ideas fit in the framework
of QCD, as discussed earlier 
\refnote{\cite{po}}. 
We then consider the 
$1+1$-dimensional
$O(2)$ and $O(3)$ nonlinear sigma models and show that no
gap exists in the former at weak coupling. In the $O(3)$ model
a new kind of strong/weak-coupling duality is realized. We
then briefly outline our proposals for understanding the spectrum.

\section{THE YANG-MILLS METRIC}

\indent In the last few decades, there have been many
serious attempts to understand

\vfill

\newpage

\topmargin 10mm

\noindent 
wave functionals on orbit space by
isolating a 
fundamental region (the interior of 
the Gribov horizon) 
\refnote{\cite{2}}. Instead we examine this space
using ``automorphic functions", i.e. gauge-invariant
wave functionals \refnote{\cite{po,lee}}. An interesting approach along
similar lines for $2+1$-dimensional gauge theories is that
of reference 4.

The Hamiltonian
of the $D+1$-dimensional $SU(N)$ Yang-Mills theory in $A_{0}=0$ (temporal)
gauge 
\begin{eqnarray} 
H= \int_{M}\,  d^{D}x\; 
[-\frac{e^{2}_{0}}{2} \,tr\,\,\frac{\delta^{2}}{\delta\, A_{j}(x)^{2}}
+ \frac{1}{4e^{2}_{0}}  \,tr\,F_{j\,k}(x)^{2}]\;. \nonumber
\end{eqnarray}  
The allowed 
wave functionals $\Psi$ satisfy the condition that if $A$ and $B$
are physically equivalent (same up to gauge 
transformations of C-S number zero)
$\Psi[A]=\Psi[B]$.

This will be reformulated as a particle on a certain 
infinite-dimensional curved space on which there is a 
height function, namely the
potential energy. We are interested ultimately in how geodesic
motion (strong couping) is influenced by this height function
(potential topography).

The lattice discussion will be used
here to introduce the Yang-Mills metric. A 
more 
heuristic motivation following Feynman \refnote{\cite{feynman}} 
was used in reference 1.

Consider a lattice gauge theory with $D$ space and $1$ time
dimension. Label discrete points in Euclidean
space-time by $x$ and $t$, respectively. Let $\{U(t)\}$ denote the set
of lattice gauge fields (in the fundamental representation
of $SU(N)$) on links pointing in space directions 
at a particular time . Split the action $S$ 
into a space-time plaquette term $S_{st}$
and a space-space plaquette term $S_{ss}$, i.e. 
$S=S_{st}+S_{ss}$, where
\begin{eqnarray} 
S_{st}=-\sum_{t} \frac{1}{2e_{0}^{2}}{\Re} [\{U(t)\}, \{U(t+1)\}]^{2} 
\;,\;\;
S_{ss}=-\sum_{t} \frac{1}{4e_{0}^{2}}{\ell}[\{U(t)\}] \;, \nonumber
\end{eqnarray}  
where
\begin{eqnarray} 
{\Re} [\{U(t)\}, \{U(t+1)\}]^{2} 
= \frac{1}{2}\sum_{x,i} 
tr [U_{i}(x,t)U_{0}(x+{\hat i},t)U_{i}(x+{\hat t},t)^{\dagger}
U_{0}(x,t)^{\dagger}]+c.c.\;, \nonumber
\end{eqnarray}  
and 
\begin{eqnarray} 
{\ell}[\{U(t)\}] =\sum_{x, i<j}
tr [U_{i}(x,t)U_{j}(x+{\hat i},t)U_{i}(x+{\hat j},t)^{\dagger}
U_{j}(x,t)^{\dagger}]+c.c.\;. \nonumber
\end{eqnarray}  
Now let's try to integrate out the links pointing in time 
direction, $U_{0}(x,t)$. As a first
approximation to doing this, we can just solve their equation of 
motion. This says that 
${\Re} [\{U(t)\}, \{U(t+1)\}]^{2}$ is minimized with respect to these
degrees
of freedom. If we integrate them out explicitly the result is a product of a 
Bessel functions. Near the maximum of this product it has the form
$\exp-\sum_{t}\rho [\{U(t)\}, \{U(t+1)\}]$, where $\rho$ is the 
{\it absolute} (not local) minimum\begin{eqnarray} 
\rho [\{U(t)\}, \{U(t+1)\}] 
=\min_{\{U_{0}(t)\}} {\Re} [\{U(t)\}, \{U(t+1)\}] \;. \nonumber
\end{eqnarray}  
The quantity $\rho [\{U(t)\}, \{U(t+1)\}]$ can be 
shown to be a metric on the space $\{U(t)\}$ modulo gauge 
transformations. Thus the kinetic term in the action
by itself describes Brownian motion in this space.

In the continuum the
space of connections $\cal U$ is defined
to contain only those gauge fields which are Lebesgue
measurable, and square-integrable \refnote{\cite{po}}. No distinction 
is made between 
gauge fields which are the
same almost everywhere (${\cal U}$ is a Hilbert space). Gauge 
transformations are $SU(N)$ valued functions $g(x)$
which
are differentiable and for which 
$ig^{-1} \partial g \in {\cal U}$. Any element of ${\cal U}$ is mapped 
into another element of ${\cal U}$ by such a gauge transformation.

The equivalence classes must actually be made
larger in order to obtain a metric space ${\cal M}_{D}$. Two vectors in 
$\cal U$ with representatives $A$ and $B$
will be said to be gauge-equivalent if there is a sequence of 
gauge transformations $g_{1}$, $g_{2}$,..., such
that
\begin{eqnarray} 
B=\lim_{n} A^{g_{n}} \nonumber
\end{eqnarray}  
in the usual metric of the Hilbert space ${\cal U}$.

Let $\alpha$ and $\beta$ be two physical
configurations, with $A$ a representative 
of $\alpha$ and $B$ a representative of $\beta$. The
distance function \refnote{\cite{feynman,atiyah}} 
is defined
by
\begin{eqnarray} 
\rho [\alpha, \beta]=  \inf_{h,f} \{ {\sqrt{ 
\frac{1}{2}\;\int_{M} d^{D}x \;tr \;[A^{h}(x)-B^{f}(x)]^{2}
}}\} \;. \nonumber
\end{eqnarray}  
This is just the continuum version of the lattice metric
defined above. The 
function $\rho [\cdot, \cdot]$ was shown to give a complete
metric space on
equivalence classes of gauge connections ${\cal M}_{D}$.

There is a local metric on the space of connections. This
turns out to be essentially 
that discussed some time ago 
\refnote{\cite{math1, math2}}. The Laplacian
actually contains several terms not found by these 
authors \refnote{\cite{po, schwinger}}. The geodesics 
of the space can be proved to be those conjectured by Babelon and 
Viallet \refnote{\cite{math2}}. With the metric
tensor defined properly, there are no non-generic points as 
had been claimed 
(the orbit space is complete). We believe the problem some people
found
with non-generic points is related to the fact that they
worked with connections in Sobolev space rather than
those in ${\cal U}$ in which case there 
is no longer completenss using
the metric $\rho[\cdot, \cdot]$.

The square of the infinitesmal distance in orbit space ${\cal M}_{D}$
due to a 
small displacement $\delta A$ in $\cal U$ 
is
\begin{eqnarray} 
d\rho ^{2}
=[\int_{M} d^{D}x \;\sum_{j=1}^{D} \;\sum_{a=1}^{N^{2}-1}]
         \;[\int_{M} d^{D}y \;\sum_{k=1}^{D}\;
         \sum_{b=1}^{N^{2}-1}]
\;G_{(x, j, a) \;(y, k, b)}\; \delta A^{a}_{j}(x) 
\delta A^{b}_{k}(y)\;, \nonumber
\end{eqnarray}  
where the metric tensor is
\begin{eqnarray} 
G_{(x, j, a) \;(y, k, b)}=\delta_{j\;k} \delta_{a\;b} \delta^{D}(x-y)
-({\cal D}_{j}\; {\cal P}\frac{1}{{\cal D}^{2}} \;{\cal D}_{k})_{a\;b}\;
\delta^{D}(x-y)\;, \nonumber
\end{eqnarray}  
and where in the Green's function ${\cal P}\frac{1}{{\cal D}^{2}}$, the
principle value projects out the zero
modes of ${\cal D}^{2}$. Reducible connections 
\refnote{\cite{fuchs}} 
are a set of measure zero.

This metric tensor has zero eigenvalues; the dimension
of the coordinate space $\cal U$ is larger than the dimension
of the orbit space. One must either gauge fix 
(and deal with the Gribov problem by prescribing a fundamental
domain) or develop methods for Riemannian geometry for metric
tensors with zero modes. We follow the latter path. The 
Laplacian was first found by Schwinger on the basis of relativistic
invariance and further discussed by 
Gawedzki \refnote{\cite{schwinger}}. It was
constructed in reference 1 using 
a theory of tensors 
applicable when the dimension of coordinate
space is greater than the dimension of the manifold. It
is 
\begin{eqnarray} 
\Delta \Psi[\alpha] 
&=& - \partial_{Y}(G^{Y\,U}\partial_{U} \Psi[\alpha]) +
(\partial_{Z} G^{Z}_{Y})G^{Y\,U}\partial_{U} \Psi[\alpha]\;, \nonumber
\end{eqnarray}  
where capital Roman letters denote ``indices" $X=(x, j, a)$ and 
$\partial_{X}=\frac{\delta}{\delta A^{a}_{i}(x)}$.

The Yang-Mills Hamiltonian is
\begin{eqnarray} 
H= \frac{e^{2}}{2} \Delta+\frac{e^{2}{\cal R}}{12}
+ \int_{M}\,  d^{D}x\; 
\frac{1}{4e^{2}}  \,tr\,F_{j\,k}(x)^{2}\;, \nonumber
\end{eqnarray}  
where $\cal R$ denotes the ultraviolet divergent 
scalar 
curvature.

\section{STRUCTURE OF THE YANG-MILLS 
POTENTIAL ENERGY - RIVER VALLEYS AND GLUONS}

\indent The metric properties of the manifold ${\cal M}_{D}$ of 
configurations determine the spectrum of the kinetic term of the 
Hamiltonian. To understand the spectrum at weak coupling, the
potential or magnetic
energy must be examined.

A natural starting point is to try make a relief map
of magnetic energy on ${\cal M}_{D}$; in other words to investigate 
the magnetic topography 
\refnote{\cite{po}}.

Suppose that the manifold of physical
space is very large. Make 
a rescaling of the coordinates and the connection $A\in {\cal U}$
by a real
factor $s$:
\begin{eqnarray}                     
A_{j}(x) \longrightarrow sA_{j}(sx) \;. \nonumber
\end{eqnarray}  
A gauge-transformed connection $A^{h}$ will 
transform the same way under a rescaling, provided
$h(x)$ is redefined by
\begin{eqnarray} 
h(x) \longrightarrow h(sx) \;. \nonumber
\end{eqnarray}  
The distance of the point of orbit space $\alpha$ from an
equivalence class of pure gauges, $\alpha_{0}$, transforms
as
\begin{eqnarray} 
\rho[\alpha, \alpha_{0}] \longrightarrow s^{\frac{2-D}{2}} 
\rho[\alpha, \alpha_{0}]\;. \nonumber
\end{eqnarray}

Let $A\in {\cal U}$ be a particular configuration
of finite potential energy, for
which the magnetic field $F_{jk}(x)$  
decays rapidly to zero for $x$ outside some
finite bounded region, which
will be called the {\bf domain} of the 
magnetic field. By changing the
size of the domain and the magnitude
of the magnetic field, the distance from some given
pure gauge can be made arbitrarily small (except when 
regularization effects become important) or large (except when 
volume effects become important).

The potential energy
\begin{eqnarray} 
U[\alpha]=\int_{M}\,  d^{D}x\; \frac{1}{4e^{2}}  \,tr\,F_{j\,k}^{2}(x) \;, 
\nonumber
\end{eqnarray}  
transforms as
\begin{eqnarray}                                               
U[\alpha] \longrightarrow s^{4-D} U[\alpha]  \; \nonumber
\end{eqnarray}  
and so for $D>2$
\begin{eqnarray} 
U \sim \rho^{-2\frac{4-D}{D-2}} \;. \nonumber
\end{eqnarray}

For $2<D<4$ the exponent is negative. Thus
it is possible to have arbitrarily large $U$ for arbitrarily
small $\rho$.

For Abelian gauge theories, other rescalings
can be considered;
\begin{eqnarray} 
A_{j}(x) \longrightarrow s^{\phi}A_{j}(sx) \;, \nonumber
\end{eqnarray}  
where ${\phi}$ is any real number. By 
choosing $\phi$ satisfying
$\frac{D-2}{2} < \phi < \frac{D}{2}$
it is always possible
to make the potential energy arbitrarily small for small $s$, no matter
what the dimension. The domain can be made large, while the field 
strength is made small; a quantum wave functional $\Phi[\alpha]$
whose amplitude is largest near this configuration is a 
long-wavelength 
photon. This quantum state must
be orthogonal to the vacuum $\Psi_{0}[\alpha]$ because at least
one of the two wave functionals is 
zero at any point in orbit
space. This is
why non-compact
electrodynamics 
has no mass gap in any dimension. Our analysis
seems to indicate that the same is
true
for Yang-Mills theory for dimension between $2+1$ and $4+1$. 

Figure 1. illustrates the situation. Orbit space
contains regions, which we call {\bf river valleys} in which the potential 
energy 
vanishes in the thermodynamic limit. The configurations in the
river valleys are not pure-gauge 
configurations
$\alpha_{0}$. The river valleys are preserved under scale transformations
and are therefore finite-dimensional. In 
perturbation theory only the region
near $\alpha_{0}$ is explored. Perturbative gluons are oscillations along 
straight line extrapolations of these curves.

\begin{figure}[t]
\centering
\epsfxsize=6.0cm\epsfbox{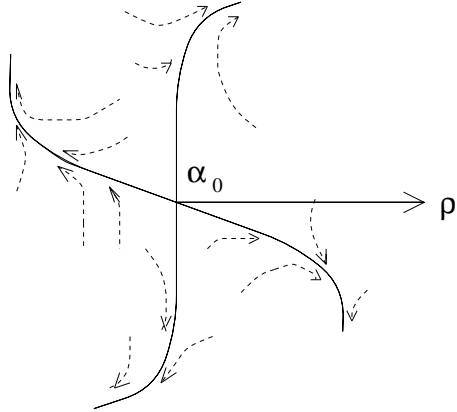}

\caption[l]{
The topography of the Yang-Mills theory in four
space-time dimensions. The dashed 
arrows represent directions of decreasing potential energy
along scale transformations. The radius $\rho$ 
is the distance from the zero potential configuration $\alpha_{0}$. The solid
curves (river valleys) are where the potential
energy
vanishes in the thermodynamic limit (only a finite subset is depicted).}
\end{figure}

\section{OUR STRATEGY}

\indent In the light of the above discussion, how could the spectrum
of QCD possibly contain
anything other that massless particles?

\noindent
{\em Answer}: The regions of small potential energy
could have large electric
(kinetic) energy by the uncertainty principle. The zero-point
energy of the modes transverse to the river valleys must
be added to the
potential. In
this
way, the first excited state could have a finite gap
above the ground state. There are
examples of quantum-mechanical systems, including 
models motivated by Yang-Mills theory \refnote{\cite{savvidy}}, for which 
this is true \refnote{\cite{simon}}.

We should view the position along
each river valley as a collective coordinate. We then integrate
out all degrees of freedom except this coordinate. The resulting
quantum-mechanical Hamiltonian will have eigenstates which
correspond to the eigenstates of the field-theory
Hamiltonian with zero total
momentum. To insure consistency of the collective-coordinate
approximation, we will must consider only small fluctuations around
river valleys.

\section{THE $O(n)$ NONLINEAR SIGMA MODEL}

\indent A great deal is known about the $1+1$-dimensional sigma
models. The phase transition in the $O(2)$ model is
well understood \refnote{\cite{kt}}. By virtue of its integrability 
\refnote{\cite{int}} the $S$-matrix \refnote{\cite{zam}} and spectrum 
\refnote{\cite{wiegmann}} of the
$O(3)$ model are also known. Unfortunately neither these methods
nor the $1/n$-expansion \refnote{\cite{stanley}}
extend to gauge theories in higher dimensions. 

This model will first be considered in $D+1$ 
dimensions. Later we will specialize to
$D=1$. The field $s(x)$ (we are fixing time) 
with $x$ on a $D$-dimensional lattice
is a real $n$-tuple with $s^{T}(x) s(x)=1$.
The Hamiltonian is
\begin{eqnarray} 
H=\frac{e_{0}}{2}\sum_{x} L(x)^{T}L(x)
-\frac{1}{2e_{0}}\sum_{<x,x^{\prime}>} s(x)^{T}s(x^{\prime})\;. \nonumber
\end{eqnarray}  
The fields $s$ lie in equivalence classes:
\begin{eqnarray} 
\psi=\{Rs;\;R\in O(n)\}. \nonumber
\end{eqnarray}

The definition of these equivalence
classes isn't yet obviously 
right. Unlike the case of Yang-Mills theory, the 
equivalence class contains physically {\em different}
configurations. We will worry about this issue later.

A natural metric on equivalence classes $\psi$, $\phi$
with $s\in \psi$, $f\in \phi$ is \refnote{\cite{feynman}}
\begin{eqnarray} 
\rho[\phi, \psi]^{2}= \inf_{R\in O(n)} \sum_{x} 
[Rf(x)-s(x)]^{T} [Rf(x)-s(x)] \;. \nonumber
\end{eqnarray}

In the continuum this goes over to (up to factors of the
lattice spacing)
\begin{eqnarray} 
\rho[\phi, \psi]^{2}&=&\inf_{R\in O(n)} \int d^{D}x \; 
[Rf(x)-s(x)]^{T} [Rf(x)-s(x)] \\ \nonumber
&=&V-tr{\sqrt {M^{T}M}}\;, \nonumber
\end{eqnarray}  
where $V$ is now the volume of the space manifold and 
\begin{eqnarray} 
M_{jk}=\int d^{D}x\; f_{j}(x)s_{k}(x) \;. \nonumber
\end{eqnarray}  
However, we no longer strictly have a metric without making certain 
restrictions on allowed spin configurations. Without such
restrictions, different configurations are separated by
a distance zero. This is a minor difficulty
and will not trouble us.

\section{THE SIGMA-MODEL
RIVER VALLEYS}

\indent Let's denote the  ``pure gauge" configuration 
containg
constant $s(x)$ by $\psi_{0}$.

Consider now the following problem for $D=1$. For
fixed $\rho[\psi, \psi_{0}]$ minimize the the potential
energy 
\begin{eqnarray} 
U(\psi)=\int_{0}^{L} \;dx\;
s^{\prime}(x)^{2}\;,
\;\;s\in\psi\;.\nonumber
\end{eqnarray}  
subject to Neumann boundary conditions 
$s^{\prime}(0)= s^{\prime}(L)=0$. Let's parametrize $s(x)$ using angles 
$\theta_{1}(x)$,..., $\theta_{n-1}(x)$, by $s_{1}(x)
=\sin\theta_{1}(x)...\,\sin\theta_{n-1}(x)$,
..., $s_{n}=\cos\theta_{1}(x)$, 
in the standard way.

The solution for the minima of $U(\psi)$ for fixed $\rho=\rho[\psi,\psi_{0}]$ 
(distance from the origin=pure gauge) is
similar to that of a pendulum. We find that up to global 
rotations $R$ there are minima
labeled by an integer $N=1$, $2$, ...
\begin{eqnarray} 
\theta_{1}(x)
=\pm \alpha_{N}(x,k)= \pm 2\sin^{-1} k\; sn(\frac{2NK}{L}x-K) \;,  \nonumber
\end{eqnarray}  
\begin{eqnarray} 
\theta_{2}(x)=\cdot \cdot \cdot =\theta_{n-1}(x)=0\;,\nonumber
\end{eqnarray}  
where $0\le k \le 1$ is the modulus of the elliptic function $sn(u)$ 
and $K=K(k)=sn^{-1}(1)$
is
the usual complete elliptic integral. The
river valleys are nicely parametrized by the modulus $k$ as shown in figure
2.

We find that
\begin{eqnarray} 
\rho (k)^{2}= \left\{ \begin{array}{cc}  
2L(1-\frac{E}{K})\;, & \;0\le k \le k^{*}\approx 0.82 \\
2L\frac{E}{K} \;\;\;,    & \;k^{*} \le k \le 1
\end{array} 
\right. 
\;,
\end{eqnarray}  
where $E=E(k)=\int_{0}^{1} dn^{2}(u) \,du$ is another standard elliptic
integral. This function rises smoothly from $0$ to $L$ as $k$ goes 
from $0$ to $k^{*}$, then
falls off to zero again as $k\rightarrow 1$. In fact, on the lattice the 
$k=1$ solution
is unphysical, because this 
solution has discontinuities in the continuum. Actually 
$k\le k_{max}\approx 1$
because of the regulator. A configuration along a river valley 
is maximally
far from the origin at $k=k^{*}$.

The potential
energy function is
\begin{eqnarray} 
U(k)=\frac{32N^{2}K}{L} [E-(1-k^{2})K] \;. \nonumber
\end{eqnarray}  
For fixed volume $L$ this diverges at $k=1$, but, as mentioned 
earlier, this divergence
is regularized by a lattice (or some other ultraviolet cut-off).

In the infinite volume limit for $k<k^{*}$, the potential is a 
constant; but the one-dimensional
domain over which this is so has an infinite length (=$L$). If we 
view $k$ as a collective
variable, and ignore fluctuations in other degrees of freedom 
the gap is $O(\frac{1}{L})$.

\begin{figure}[t]
\centering
\epsfxsize=6.0cm\epsfbox{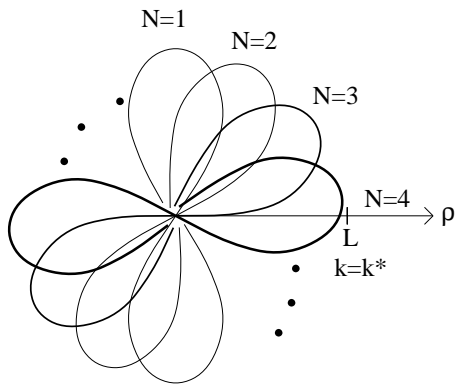}
\caption[l]{The river valleys for the $O(n)$ sigma model. The 
potential energy is nearly constant
along the solid curves. As before, a straight line 
extrapolation along tangent
vectors at the origin gives the spin wave approximation.}\end{figure}

We note that the river valleys are not straight lines in 
configuration space. Their 
tangent vectors at $k$
are 
\begin{eqnarray} 
\beta_{N}(x,k)=\frac{\partial \alpha_{N}(x,k)}{\partial k}= 
\frac{2sn(u)dn(u)-Z(u)cn(u)}{1-k^{2}} \;, \nonumber
\end{eqnarray}  
where $u=2NKx/L-K$ and $Z(u)$ is the Jacobi zeta 
function. The inner
product of $\beta_{N}$ and its derivative with respect 
to $k$ is not zero; this means that 
the river valleys are curved. The tangent vector does not
have unit length in our collective coordinates. It is convenient to
define the unit tangent vector
${\hat \beta}_{N}(x,k)=\beta_{N}(x,k)/{\sqrt {{\int_{0}^{L} 
\beta_{N}(y,k)^{2}\,dy}}}$.

\section{COLLECTIVE COORDINATES}

\indent Up to now we've ignored the fact that a system with global 
symmetry has states which transform
as some representation of that symmetry (For Yang-Mills theories, we have no
such problem). For example if $n=2$ our river valleys are not one-dimensional,
but
two-dimensional surfaces parametrized by 
$\theta_{0}$ as well as $k$: $\theta(x)=\theta_{0}\pm\alpha_{N}(x,k)$. In
fact the river valleys of the $O(n)$ sigma model 
are really $n-1$-dimensional manifolds. However
this consideration is irrelevant if we are asking 
for only certain information. The degree of freedom
corresponding to $\theta_{0}$ is the very longest 
wavelength Goldstone mode. We
can remove this mode if we are interested in mass 
spectra only and don't care about degeneracies
of our states. For example, this
can be done in the $O(3)$ model by adding a term $-\int dx dt\; 
\frac{\lambda}{V} s_{1}^{2}/(1-s_{2}^{2})$, where $V$ is the space-time 
volume. Such a term is of no consequence in the thermodynamic limit but 
clearly 
keeps the river valleys one-dimensional, rather than three-dimensional
manifolds. This won't
matter as long as we aren't interested 
in the transformation properties or non-accidental degeneracies
of our states. 


The collective-coordinate representation \refnote{\cite{gervais}} of 
$\theta_{1}(x)$ is 
\begin{eqnarray} 
\theta_{1}(x,t)=\alpha_{N}(x,k(t))+\sum_{a=1}^{\infty} w_{a}(t)
T^{a}_{N}(x,k(t))\;,   \nonumber
\end{eqnarray}  
where we have now taken the range of $k(t)$ to be $-k_{max}<k(t)<k_{max}$
and $\alpha_{N}(x,-k) \equiv -\alpha_{N}(x,k)$. The family
of functions $T^{a}_{N}(x,k)$ satisfy
\begin{eqnarray} 
\int_{0}^{L} T^{a}_{N}(x,k)T^{b}_{N}(x,k)\; dx=\delta^{ab}\;,\;\;  
\int_{0}^{L} T^{a}_{N}(x,k) {\hat \beta}_{N}(x,k)\; dx=0\;,  \nonumber
\end{eqnarray}  
and $T^{a\;\prime}(0)=T^{a\;\prime}(L)=0$. 

\begin{figure}[t]
\centering
\epsfxsize=6.0cm\epsfbox{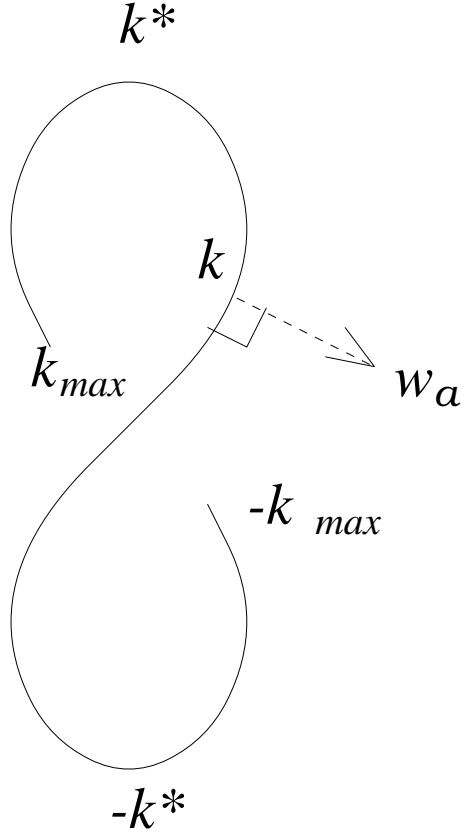}
\caption[l]{A depiction of the collective 
coordinate representation. The degree
of freedom $k(t)$ parametrizes the river 
valley, while the $w_{a}(t)$'s parametrize
normal displacements}\end{figure}

Physically, the collective coordinate $k$ is the parameter along 
the $N^{th}$ river valley, while the $w_{a}(x)$'s are coordinates of
dispacements normal to the valley at the point described by $k$.

In order to proceed further, it is necessary to compute the Jacobian
$\Vert J \Vert$ of the transformation to 
the collective coordinate system. The details of 
this computation will be presented elsewhere. The answer turns out 
to be 
\begin{eqnarray}
\Vert J \Vert= 
{\sqrt {\int_{0}^{L}\beta_{N}(x,k)^{2}\,dx}}
-\sum_{a=1}^{\infty} w_{a} \int_{0}^{L}
\frac{\partial {\hat \beta}_{N}(x,k)}{\partial k} 
T^{a}(x,k)\,dx\;.\nonumber
\end{eqnarray}

\section{NO GAP FOR THE 
$O(2)$ SIGMA MODEL}

\indent The behavior of the $O(2)$ sigma model at weak coupling is generally
regarded as obvious \refnote{\cite{kt}}. From our 
perspective however, this model is nearly as hard to understand 
as all the others.

The functional measure on $\theta_{1}=\theta$ is fairly simple. The fact that
it is compact is responsible for vortices which are 
transitions between river valleys
We will ignore these transitions for the following 
reason. A vortex can be regarded as a process
through which a configuration 
evolves in time; the initial and final configurations are in the
same equivalence class. Thus it is a loop 
in the metric space. The length of this loop
can be computed and 
diverges in the thermodynamic limit. In the spirit of our approach, which is
to consider only small 
fluctuations near river valleys, vortices can therefore be ignored. The 
details of this elementary
calculation will be presented elsewhere. 

The only nontrivial factor in the measure
is the Jacobian to collective coordinates. The path integral 
\begin{eqnarray} 
Z=\int {\cal D} \theta(x,t) \exp-\int\, dt\,\int_{0}^{L}\,dx\, 
\frac{1}{2e_{0}}(\partial_{t}\theta^{2}+\partial_{x}\theta^{2})  \nonumber
\end{eqnarray}  
can be expanded to quadratic order
in $w_{a}$:
\begin{eqnarray} 
Z\approx \int {\cal D} k(t) 
{\cal D}w_{a}(t) [\prod_{t}\Vert J(k(t), w_{a}(t))\Vert]  
\; \exp -\frac{1}{2e_{0}}\int dt \{
                 [\int_{0}^{L}\beta_{N}(x,k)^{2}\,dx]\; {\dot k}^{2}         
\nonumber 
\end{eqnarray}  
\begin{eqnarray} 
+  \sum_{a} {\dot w}^{2}_{a} + \sum_{ab} w_{a} \Omega^{ab}(k) w_{b}    
+source\; terms\; for\; w_{a} \}\;, \nonumber
\end{eqnarray}  
where the matrix $\Omega(k)$ is the projection of the 
operator $-\frac{d^{2}}{dx^{2}}$ onto the subspace of
normalizable functions satisfying Neumann boundary conditions 
and which are orthogonal to $\beta_{n}$. In other
words 
\begin{eqnarray} 
\Omega^{ab}(k)=\int_{0}^{L} 
\frac{dT^{a}(x)}{dx} 
\frac{dT^{b}(x)}{dx} 
\,dx\;. \nonumber
\end{eqnarray}  
The source
term can be shown to be unimportant in the limit of infinite time evolution.

Only the first term in $\Vert J \Vert$ should be included to one 
loop. This factor cancels
out if a change of variable is made from $k(t)$ to the arc-length 
parameter $\gamma(t)$ defined
by 
\begin{eqnarray} 
\frac{d \gamma}{dk} = {\sqrt{\int_{0}^{L}\beta_{N}(x,k)^{2}\,dx}} 
\;.           \nonumber 
\end{eqnarray}  
We will write $k(t)$ as $k(\gamma(t))$ henceforth. For small 
$k$, $\gamma \approx \rho \approx k{\sqrt L}$.

The zero-point-energy contribution from the $w_{a}$'s is 
simply half the sum of eigenfrequencies. The
square of any one of these eigenfrequencies is an 
eigenvalue of $\Omega(k(\gamma))$. After
integrating out these modes, we are left with
\begin{eqnarray} 
Z=\int {\cal D} \gamma(t) \exp - \int \;dt\; 
[\;\frac{{\dot \gamma}(t)^{2}}{2e_{0}} 
+ \frac{1}{2} tr {\sqrt{\Omega(k(\gamma(t)))}}
- \frac{1}{2} tr {\sqrt{\Omega(0)}}
\;\;] \;. \nonumber
\end{eqnarray}

Computing the eigenvalues of $\Omega(k)$ is done in the following
way. The operator
\begin{eqnarray} 
\omega(k)= 
(1-\vert {\hat \beta}_{N}><{\hat \beta}_{N} \vert)\; 
(-\frac{d^{2}}{dx^{2}})\;
(1-\vert {\hat \beta}_{N}><{\hat \beta}_{N} \vert) \nonumber
\end{eqnarray}  
on the Hilbert space on $[0,L]$ with Neumann boundary
conditions has the same eigenvalues as $\Omega(k)$. Here we use
standard Dirac bra-ket notation for Hilbert space vectors. The 
eigenvalue
problem for this operator is straightforward (we only present a
quick and dirty derivation of the answer here).
 
Let $G(\lambda)$ be the inverse of $-d^{2}/dx^{2}-\lambda$ on the
Hilbert space. The operator $G(\lambda)$ isn't regular, since it has
poles if $\lambda=n^{2}\pi^{2}/L^{2}$, but
$G(\lambda) \sin\lambda $ is regular. Consider
$G(\lambda) \sin\lambda \vert {\hat \beta}_{N}>$. If this is
is orthogonal to $\vert {\hat \beta}_{N}>$, it must be an
eigenvector of $\omega(k)$ with eigenvalue $\lambda$. By 
doing the analysis more carefully one can show
that all the eigenvectors are
of this form. The condition that $\lambda$ be an eigenvalue of $\Omega(k)$
is therefore 
$\sin\lambda <{\hat \beta}_{N}\vert G(\lambda)  \vert {\hat \beta}_{N}>=0$ or
\begin{eqnarray} 
\int_{0}^{L} dx\,\int_{0}^{L-x} dy\,\, \cos \lambda x           
\cos \lambda y \;
{\hat \beta}_{N}(x)
{\hat \beta}_{N}(y)=0\;. \nonumber
\end{eqnarray}  
This function can be computed numerically, the zeros can be found, and the sum of the
square roots obtained. A 
graph of the potential versus $k$ (not $\gamma$) has
vanishing slope at the
origin and rises significantly only for $k\approx k^{*}$. One can
therefore conclude that the gap to the first excited
state is of order $1/L$.

Even without doing a very explicit calculation one can see that gap is
impossible by a simple scaling argument. Each eigenvalue is directly proportional
to $1/L^{2}$ and the sum of square roots of eigenvalues is
finite (after making the subtraction at $k=0$. Therefore, this sum must 
have the form
\begin{eqnarray} 
\frac{1}{2}\,tr {\sqrt{\Omega(k(\gamma))}} = \frac{1}{L} f(k(\gamma)) \;.
\nonumber
\end{eqnarray}  
The only way any nontrivial
$\gamma$ dependence could emerge in the 
thermodynamic limit is if 
there is a term in
$f(k)$ proportional to $1/k^{2}$, for small $k$. But then the result
at large $L$ would be 
\begin{eqnarray} 
\frac{1}{2}\, tr {\sqrt{\Omega(k(\gamma))}} = 
\frac{C}{\gamma^{2}} \;,\nonumber
\end{eqnarray}  
where $C$ is a constant for $\gamma < \sqrt L$. But this
is a Calogero potential which has a continuous
spectrum (there is no harmonic oscillator term).

\section{THE $O(3)$ CASE AND THE LAM\'E EQUATION}

\indent The $O(3)$ model has two angles 
$\theta_{1}=\theta$ and $\theta_{2}=\phi$. Let us consider
the lattice path integral 
\begin{eqnarray} 
Z=[\prod_{x,t} 
\int_{-\pi/2}^{\pi/2} d \theta(x,t) 
\;\vert \sin\theta(x,t) \vert \;] 
[\prod_{x,t} 
\int_{-\pi}^{\pi}  d\phi(x,t)]  \nonumber
\end{eqnarray}  
\begin{eqnarray} 
\times \;\exp-\sum_{x,t}\, \frac{1}{2e_{0}} \{ 
[1-\cos (\theta(x,t+a)-\theta(x,t))]
+[1-\cos (\theta(x+a,t)-\theta(x,t))] \nonumber
\end{eqnarray}  
\begin{eqnarray} 
+\sin^{2} \frac{\theta(x,t+a)+\theta(x,t)}{2}
\;[1-\cos (\phi(x,t+a)-\phi(x,t))] \nonumber
\end{eqnarray}  
\begin{eqnarray} 
+
\sin^{2} \frac{\theta(x+a,t)+\theta(x,t)}{2}
\;[1-\cos (\phi(x+a,t)-\phi(x,t))]
\}  \;. \label{4}
\end{eqnarray}  
Here $x$ and $t$ are discrete, namely 
integers times the lattice spacing $a$ (which
is assumed to be much smaller than $L/N$). To 
calculate the zero-point energy of the fluctuations to one loop, the 
quantities $\sin^{2} (\theta(x+a,t)+\theta(x,t))/2$ 
in the action can be replaced by
$\sin^{2} \alpha_{N}$. Similarly the 
measure factor $\prod_{x,t} \sin \vert \theta(x,t) \vert$ can
be replaced by 
$\prod_{x,t} \sin \vert \alpha_{N}(x,k(t)) \vert$. Notice 
that $\sin^{2} \alpha_{N}$ vanishes
at $x_{j}
=\frac{(2j-1)L}{2N}$, $j=1$,...,$N$. This 
means that Neumann boundary conditions $\partial_{x}\phi=0$
arise at these points; furthermore $\phi$ can be 
discontinuous at $x_{j}$. The 
degrees of freedom in the field $\phi$ do not
communicate with one another
across the line $x=x_{j}$. This breaking 
of the part of the action depending on $\phi$ into
independent pieces in strips 
separated by the $x_{j}$ is an artifact of the one-loop
approximation. At higher loops, we 
can no longer assume the coefficients of the $\phi$ lattice
derivatives vanish at these points.

As in the $O(2)$ case, vortex configurations of the $\theta$ field are
of no importance at weak coupling. The contribution the fluctuations of
this field give
when integrated out is the same as before.

Consider next the integration over $\phi$. If 
$\gamma << L$ (i.e. $\vert k \vert <<1$), then 
the coefficient of the $\phi$ lattice derivatives, namely
$\sin^{2}\alpha_{N}(x,k(t))$ is small over most of 
spacetime for small $k$ and large over most
of spacetime for large $k$. This function is essentially a 
slowly-varying 
inverse coupling constant for $\phi$. We therefore expect
$\phi$-field vortices
to be important, and we cannot treat the integral as a Gaussian
in $\phi$. However, let us come back to this point a little later and see what
happens if the Gaussian approximation is used for the $\phi$ integration.

If we assume that $k(t)$ is slowly varying and ignore 
its time derivatives, we can absorb the measure factor
$[\prod_{x,t} \sin\alpha_{N}(x,k(t))]$ into ${\cal D}\phi(x,t)$, by
defining $\Phi=\phi \sin\alpha_{N}$ and take the 
continuum
limit, obtaining 
\begin{eqnarray} 
Z&=&\int {\cal D} \theta(x,t) 
    \int {\cal D} \Phi(x,t) 
    \exp-\int\, dt\, 
    \int_{0}^{L}\,dx\,
    \frac{1}{2e_{0}} \{ \partial_{t}\theta^{2}+
     \partial_{x}\theta^{2} \nonumber \\
 &+&[\partial_{t}\Phi^{2}+
     \sin^{2}\alpha_{N}\;
      (\partial_{x}\frac{\Phi}{\sin\alpha_{N}})^{2}] \}  \nonumber \\
 &=&\int {\cal D} \theta(x,t) 
    \int {\cal D} \Phi(x,t) 
    \exp-\int\, dt\,     \int_{0}^{L}\,dx\,
    \frac{1}{2e_{0}} [\partial_{t}\theta^{2}+
     \partial_{x}\theta^{2} \\ \nonumber
 &+&\Phi\;(-\partial_{x}^{2}
    +\frac {24N^{2}K^{2}} {L^{2}}  k^{2} sn^{2}(u)-
\frac{4N^{2}K^{2}}{ L^{2}}
(1+4k^{2})\;)\;\Phi] \;,
\nonumber
\end{eqnarray}  
supplemented by Neumann 
boundary conditions $\partial_{x}\Phi=0$ at $x=0,L$ and Dirichlet
boundary conditions $\Phi=0$ at the points $x_{j}$.

After making the transformation 
from $\theta$ to $k$, $w_{1}$, $w_{2}$,... and integrating out
both the $w_{a}$'s and $\Phi$ gives
\begin{eqnarray} 
Z&=&\int {\cal D} \gamma(t) \exp - \int \;dt\; 
      [\;\frac{{\dot \gamma}(t)^{2}}{2e_{0}} 
    + \frac{1}{2} tr {\sqrt{\Omega(k)}}
    - \frac{1}{2} tr {\sqrt{\Omega(0)}}           \nonumber \\ 
 &+& \frac{2NK}{L} tr {\sqrt{{\cal L}(k)+(1+4k^{2})}}
    -\frac{N\pi}{L}  tr {\sqrt{{\cal L}(0)+1}}\;]           \label{W}      
\end{eqnarray}  
where $\Omega$ is defined as before and ${\cal L}$ 
is the Lam\'e operator
\begin{eqnarray} 
{\cal L}(k)= -\frac{d^{2}}{du^{2}}+m(m+1) k^{2} sn^{2}(u) \;, \nonumber
\end{eqnarray}  
where $m=2$ in our case.

The 
eigenvalue problem ${\cal L} \Lambda(u) = {\cal A} \Lambda(u)$ is called
the Lam\'e equation and was first solved in some generality by
Hermite (see the book by 
Whittaker and Watson \refnote{\cite{ww}}). 



A rough argument shows that the Gaussian approximation for $\phi$ does not
yield a gap between the ground state and the first excited state. Large 
eigenvalues of the Lam\'e operator are well approximated by the
eigenvalues of the Schr\"{o}dinger 
operator with zero potential. The 
fourth term in the exponent of (\ref{W}) is then an
ultraviolet-divergent expression of the form
\begin{eqnarray} 
S(\Lambda, L, k)=
\frac{A}{L} \sum_{n=1}^{\Lambda L} {\sqrt{n^{2}+f(k)}}+o(\frac{1}{L})\;, 
\nonumber 
\end{eqnarray}  
where $A$ is a constant 
and $f(k)$ is some function with no $x$-dependence. The reason the mode sum
is cut off at $\Lambda L$ is because that is the number of degrees of 
freedom in the problem (for example, on a lattice, where $\Lambda$ is the
inverse lattice spacing). For large $\Lambda L$, the sum becomes an
integral which can be evaluated to be
\begin{eqnarray} 
S(\Lambda, L, k)=\frac{Af(k)^{2}}{2L}[\sinh^{-1}\frac{\Lambda L}{f(k)}
+\frac{1}{2}\sinh(2\sinh^{-1}\frac{\Lambda L}{f(k)}]+o(\frac{1}{L})\;. 
\nonumber 
\end{eqnarray}  
For large $L$ all that remains is
$S(\Lambda,L,k)\approx \frac{A\Lambda^{2}}{2}$ which has no $k$-dependence.

\section{STRONG/WEAK-COUPLING DUALITY}

\indent Let us now look once again at (\ref{4}). We 
will make a Gaussian approximation for $\theta$, which
we know to be justified, but 
not $\phi$ (we can differentiate between $\theta$ and $\phi$ using
the arguments in section 6). In 
order to find the effective action for $\gamma$, we need to find
the contribution to the 
potential which is the free energy of the $\phi$ field with $\theta$ 
set equal to $\alpha_{N}$, i.e.
\begin{eqnarray} 
W(\gamma)=-\lim_{T\rightarrow \infty} \frac{\log Z_{\phi}(\gamma)}{T}\;,  
\nonumber 
\end{eqnarray}  
where $T$ is the time duration,
\begin{eqnarray} 
Z_{\phi}(\gamma)
&=     &[\prod_{x,t} 
        \int_{-\pi}^{\pi}  d\phi(x,t) \beta(x,\gamma)^{\frac{1}{2}}]  
\nonumber \\ &\times& \;\exp-\sum_{x,t}\, \beta(x,\gamma)
         \{[1-\cos (\phi(x,t+1)-\phi(x,t))] \\ \nonumber 
&+     &[1-\cos (\phi(x+1,t)-\phi(x,t))]
\}  \;, \label{5}
\end{eqnarray}  
and 
\begin{eqnarray} 
\beta(x,\gamma)=\frac{1}{2e_{0}}\sin^{2} \alpha_{N}(x,k(\gamma)) = 
\frac{2k(\gamma)^{2}}{e_{0}}\,sn^{2}(u,k(\gamma))\;dn^{2}(u,k(\gamma)), 
\nonumber 
\end{eqnarray}  
plays the role of inverse coupling 
constant. 


What is very striking is that no matter 
how small the coupling $e_{o}$ may be, the
effective coupling in the ``$\phi$ sector" is large for 
$\gamma=k^{2}\sqrt{L}<< \sqrt{L}$.
This is a kind of strong-coupling/weak-coupling duality. It tells
us that to study $W(\gamma)$ at weak coupling, we need a strong-coupling 
expansion. If there is a minimum of $W(\gamma)$ for
finite $\gamma$, compactness effects, i.e. {\em vortices} are
responsible. How can this be reconciled 
with the philosophy of our
approximation, namely that only configurations 
close to the river valleys may
be considered? The answer is that, unlike the case 
of the $O(2)$ sigma model, vortices
are {\em short} paths in configuration space, whose lengths are not
divergent. This point will be discussed in a
later 
publication.

We have not yet proved the existence of a gap from the ground
state to the first excited state, but it seems clear how the
proof should go. First, the strong-coupling expansion will yield
the potential $W(\gamma)$ (we have already found this). Then it must be 
checked that the gap does not disappear as $L\rightarrow \infty$. If
this is so, the spatial correlation functions must automatically
fall off exponentially; for if the wave function is localized
at small $k$, the effective 
coupling of the $\phi$-field must be strong. This is an important
check of Lorentz invariance. Finally
the dependence of the gap on $e_{0}$ must be checked for consistency
with asymptotic freedom. 


\section*{ACKNOWLEDGEMENTS}

\indent We are grateful to Paul Wiegmann for discussions. Several
years ago, Michael Aizenman proposed a scheme for 
proving exponential decay in the $O(3)$ sigma model using vortices
associated with a $O(2)$ subgroup. We thank him for describing
his ideas to us. P.O. thanks the organizers
of the workshop for the opportunity to present this work. The work
of M.K.
and P.O. was supported in part by PSC-CUNY grants, nos. 
6-67438 and 6-68460. The work
of E.M. and P.O. 
was supported in part by a CUNY Collaborative Incentive Grant, no.
991999.


\begin{numbibliography}
\bibitem{po} P. Orland, Niels Bohr preprint NBI-HE-96-35, hep-th/9607134. 
\bibitem{2} D. Zwanziger, {\it Nucl. Phys.} B209:336 (1982); 
Nucl. Phys.  B412:657 (1994); P. Koller and P. van Baal, 
{\it Ann. Phys.} 174:299 (1987); {\it Nucl. Phys.} {B302}
(1988) 1; P. van Baal, {\it Phys. Lett.} {224B} (1989) 
397; P. van Baal and N.~D. Hari-Das, {\it Nucl. Phys.} 
B385:185 (1992); P. van Baal 
and B. van den Heuvel, {\it Nucl. Phys.} B417:215 (1994); R.~E. 
Cutkosky, {\it J. Math. Phys.} 25:939 (1984); R.~E. 
Cutkosky and K. Wang, {\it Phys. Rev.} D37:3024 (1988); K. 
Fujikawa, {\it Nucl. Phys.} B468:355
(1996). 
\bibitem{lee} R. Friedberg, T.~D. Lee, Y. Pang
and H.~C. Ren, {\it Ann. Phys.} 246:381 (1996).
\bibitem{kn} D. Karabali and 
V.~P. Nair, {\it Nucl. Phys.} B464:135 (1996); {\it Phys. Lett.} B379B:141 
(1996); D. Karabali, C. Kim and V.~P. Nair, CCNY preprint CCNY-HEP-97-5, 
hep-th/9705087 (1997).
\bibitem{feynman} R.~P. Feynman, {\it Nucl. Phys.} B188:479 (1981).
\bibitem{atiyah} M.~F. Atiyah, N.~J. Hitchin and I.~M. Singer, {\it Proc.
R. Soc. Lond.} A.362:425 (1978).
\bibitem{math1} I.~M. Singer, {\it Physica Scripta} 24: 817 
(1980); P.~K. Mitter and C.~M. Viallet, {\it Commun. Math. 
Phys.} 79:457 (1981); {\it Phys. Lett.} 85B:256
(1979); M.~S. Narasimhan
and T.~R. Ramadas, {\it Commun. Math. Phys.} 67:21 (1979); M. Asorey and 
P.~K. Mitter, {\it Commun. Math. Phys.} 80:43 (1981).
\bibitem{math2} O. Babelon and C.~M. Viallet, {\it Commun. Math. 
Phys.} 81:515 (1981); {\it Phys. Lett.} 103B:45
(1981).
\bibitem{schwinger} J. Schwinger, {\it Phys. Rev.} 125:1043 
(1962); 127:324 (1962); K. Gawedzki, {\it Phys. Rev.} D26:3593 
(1982). 
\bibitem{fuchs} J. Fuchs, M.~G. Schmidt and C. Schweigert, {\it Nucl. 
Phys.} B426:107 (1994).  
\bibitem{savvidy} S.~G. Matinian, G.~K. Savvidy 
and N.~G. 
Ter-Arutunian 
Savvidy, {\it J.E.T.P. Lett.} 34:590 (1981); 53:421 (1981).
\bibitem{simon} B. Simon, {\it Ann. Phys.} 146:209 (1983).
\bibitem{kt} V.~J. Berezinskii, {\it J.E.T.P.} 32:493 (1971); J.~M. 
Kosterlitz and D.~V.  Thouless, {\it J. Phys.} C6:1181 (1973); J.~M. 
Kosterlitz, {\it J. Phys.} C7:1046 (1974);
\bibitem{int} K. Pohlmeyer, {\it Commun.
Math. Phys.} 46:207 (1976); A.~M. 
Polyakov, {\it Phys. Lett.} 72B:224 (1977); V. 
Zakharov and A. Mikhailov, {\it J.E.T.P.}
27:47 (1978); M. L\"{u}scher and 
K. Pohlmeyer, {\it Nucl. Phys.} B137:46 
(1978); E. Brezin, C.
Itzykson, J. Zinn-Justin and J.-B. Zuber, {\it Phys. Lett.} 82B:442 
(1979).
\bibitem{zam} Al.~B. Zamolodchikov and A.~B. Zamolodchikov, {\it Ann. Phys.}
120:253 (1979).
\bibitem{wiegmann} A.~M. Polyakov and P.~B. Wiegmann, {\it 
Phys. 
Lett.} 131B:121 (1983); P.~B. 
Wiegmann, {\it Phys. 
Lett.} 152B:209 (1985). 
\bibitem{stanley} H.~E. Stanley, {\it Phys. Rev.} 176:718 (1968)
\bibitem{gervais} J.-L. Gervais and B. Sakita, Nucl. Phys. B91:301 
(1975).
\bibitem{ww} E.~T.Whittaker and G.~N. Watson, ``A 
Course of Modern Analysis," Cambridge University Press, Cambridge (1927).
\end{numbibliography}
\end{document}